\documentclass[journal,twoside,web]{ieeecolor}
\usepackage{generic}
\usepackage{bm,cite,color,graphicx,epstopdf}
\usepackage{amsmath,amssymb,amsfonts,mathtools,multirow,soul,subcaption}
\usepackage{algorithmic}
\usepackage{graphicx}
\usepackage{textcomp}
\usepackage[colorlinks=true, allcolors=blue]{hyperref}
\usepackage[linesnumbered,ruled,vlined]{algorithm2e}
\DeclarePairedDelimiter\ceil{\lceil}{\rceil}
\DeclarePairedDelimiter\floor{\lfloor}{\rfloor}
\def\BibTeX{{\rm B\kern-.05em{\sc i\kern-.025em b}\kern-.08em
    T\kern-.1667em\lower.7ex\hbox{E}\kern-.125emX}}
\begin{document}
\title{A Framework of Hierarchical Attacks to Network Controllability}
\author{Yang~Lou,~\IEEEmembership{}
		Lin~Wang,~\IEEEmembership{Senior~Member,~IEEE},
		and~Guanrong~Chen,~\IEEEmembership{Life~Fellow,~IEEE}
\thanks{Y. Lou and G. Chen are with the Department of Electrical Engineering, City University of Hong Kong (e-mails: felix.lou@my.cityu.edu.hk; eegchen@cityu.edu.hk).}%
\thanks{L. Wang is with the Department of Automation, Shanghai Jiao Tong University, Shanghai 200240, China, and also with the Key Laboratory of System Control and Information Processing, Ministry of Education, Shanghai 200240, China (e-mail: wanglin@sjtu.edu.cn).}%
\thanks{This research was supported in part by the National Natural Science Foundation of China under Grant 61873167, and in part by the Natural Science Foundation of Shanghai under Grand 17ZR1445200.}%
\thanks{(\textit{Corresponding author: Lin~Wang.})}}
\maketitle

\begin{abstract}
Network controllability robustness reflects how well a networked dynamical system can maintain its controllability against destructive attacks. This paper investigates the network controllability robustness from the perspective of a malicious attack. A framework of hierarchical attack is proposed, by means of edge- or node-removal attacks. Edges (or nodes) in a target network are classified hierarchically into categories, with different priorities to attack. The category of critical edges (or nodes) has the highest priority to be selected for attack. Extensive experiments on nine synthetic networks and nine real-world networks show the effectiveness of the proposed hierarchical attack strategies for destructing the network controllability. From the protection point of view, this study suggests that the critical edges and nodes should be hidden from the attackers. This finding helps better understand the network controllability and better design robust networks.
\end{abstract}


\section{Introduction}
\label{sec:intro}

\IEEEPARstart{M}{any} real-world systems can be modeled as complex networks, which have gained growing recognition and popularity since the late 1990s, now becoming a self-contained discipline encompassing computer science, systems engineering, statistical physics, applied mathematics, and social sciences \cite{Newman2010N,Chen2014Book,Barabasi2016NS,Chen2019Book}. In practical applications, it is essential to determine whether or not a networked system can be controlled for utilization. Consequently, network controllability has become a focal research topic in network studies \cite{Liu2011N,Yuan2013NC,Posfai2013SR,Menichetti2014PRL,Motter15CHAOS,Wang2016AUTO,Liu2016RMP,Wang2017RSPTA,Wang2017SR,Zhang2017TAC,Xiang2019CSM}. Same as the classical concept for systems, \textit{controllability} here refers to the ability of a dynamical network being steered by external inputs from any initial state to any desired target state under an admissible control input within a finite duration of time.

On the other hand, random failures and malicious attacks on complex networks have become more and more frequent and severe recently \cite{Holme2002PRE,Shargel2003PRL,Schneider2011PNAS,Liu2012PO,Bashan2013NP,Xiao2014CPB}. Such failures and attacks take place in the form of node- and edge-removals, causing significant consequences to the systems such as malfunctioning or even completely crashing. For example, failures of traffic lights may cause traffic congestion in the urban transportation networks; neurological disorders may cause dysfunction or illness to humans. To resist attacks or failures, strong robustness is desirable and often necessary for a practical networked system. In different scenarios, there are different definitions and measures for network robustness \cite{Liu2017FCS}. Since theoretical analysis seems impossible for large-scale complex networks, at least in the present time, the correlation between network robustness and topological features are generally investigated empirically, taking advantages of super-computing power available today \cite{Liu2017FCS,Yamashita2019COMPSAC,Chen2019TCASII}. In this pursuit, it is worth mentioning that the development of deep learning techniques offers an efficient option for empirical studies \cite{Li2019ISBSCI,Lou2019TCYB,Fan2020NMI}.

The notion of random failures and malicious attacks on complex networks, as well as the corresponding network robustness, covers a broad range of subjects. This paper concerns with the network \textit{controllability robustness}, which refers to how well a networked dynamical system can maintain its controllability against random failures or, in particular, intentional attacks.

The issue of network robustness within different contexts regarding network topologies has been extensively investigated, and many edge- and node-removal attack strategies have been proposed to destruct the \textit{connectedness} of the networks. Generally, attack strategies can be categorized into \textit{random} and \textit{targeted} attacks. A targeted attack aims at removing an intentionally selected object (e.g., the highest-degree node or largest-betweenness edge), while a random attack do the removal randomly.

In the above studies, it is commonly assumed that necessary knowledge of the network is known and is updated after each attack. For targeted attacks, it is also assumed that the targeted object is more important than the others in maintaining the network connectedness. Commonly used measures of importance include degree centrality, betweenness centrality, neighborhood similarity \cite{Ruan2017APS}, branch weighting \cite{Vsimon2017MOE}, and structural holes \cite{Yang2020SYM}. However, ranking the importance of nodes or edges is practically intractable for large-scale networks, since most measures cannot guarantee that removing the targeted object will definitely cause a greatest effect of damage on the network.

The size of the largest connected component (LCC) is widely used as a measure for connectedness robustness \cite{Schneider2011PNAS}. It is observed that betweenness-based attacks may become less effective in the later stage of an attack process. This observation consequently leads to the effective conditional attack strategy: to remove the global highest-betweenness node only if it belongs to LCC; otherwise, to remove the local highest-betweenness node inside the LCC \cite{Nguyen2019PA}. In \cite{Nie2015PA}, degree and betweenness are used simultaneously, with predefined weights for their balance, as the measure of importance. The module-based attack strategy \cite{daCunha2015PO,Shai2015PRE} aims at attacking the nodes with inter-community edges, which are crucial to maintain the connectedness among communities. The damage-based attack \cite{Wang2014PA} uses the degree of \textit{damage} to measure the effectiveness of an attack, where the damage of an attack is defined by the change of the LCC size before and after the attack. It is also observed that the evolution process of attack and defense can enhance the network robustness \cite{Ma2016PA}, which is similar to the process of a mutual improvement of spears and shields.

\begin{figure}[htbp]
	\centering
	\includegraphics[width=.9\linewidth]{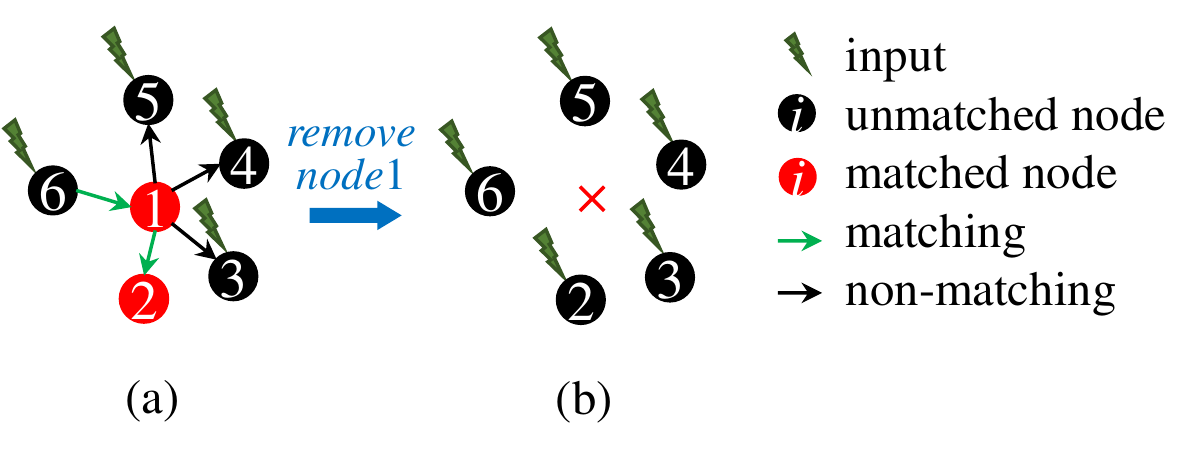}
	\caption{[color online] Example of controllability and connectedness robustness: (a) given a star-shaped network, the number of required driver nodes is $4$; its LCC is $6$; (b) after the central hub is removed, the number of required driver nodes becomes $5$; its LCC drastically dropped to $1$.}
	\label{fig:con}
\end{figure}

Although the robustness of connectedness has a certain positive correlation with the robustness of controllability on a network, they actually have very different measures and objectives, as illustrated by the simple example shown in Fig. \ref{fig:con}, where the {\it driver node} is a node to be controlled by an input so as to make the whole network become (or return to be) controllable (after the attack). This paper is concerned with the interplay of the connectedness, attack strategies, and controllability robustness of general complex networks.

Specifically, this paper focuses on the attack strategies that aim at destructing the network controllability. It is observed that removing highest-degree nodes \cite{Pu2012PA} or highly-loaded edges \cite{Nie2014PO} are more effective to degrade the network controllability than random removals. Furthermore, in \cite{Lu2016PO} it is shown that node-removals are more harmful than edge-removals to the network controllability, and that heterogeneous networks are more vulnerable than homogeneous networks. Also, it is found that for many real-world networks the betweenness-based attacks are the most destructive to the controllability \cite{Lu2016PO}. Moreover, it is reported in \cite{Sun2016PLA} that degree-based node-removal attacks cause greater damage to local-world networks \cite{Li2003PA} with a larger local-world size, while networks with smaller local-world sizes are more robust regarding both connectedness and controllability. Notably, the hierarchical structure of a directed network enables the random upstream (or downstream) attack, which removes the upstream (or downstream) node of a randomly-picked one, resulting in a more destructive attack strategy than the simple random attacks \cite{Liu2012PO}. For a directed network, edges can be categorized into three types \cite{Liu2011N}: 1) \textit{critical} edges, whose removal destroys the network controllability; 2) \textit{redundant} edges, whose removal has no influence on the controllability; and 3) \textit{ordinary} edges, whose removal will not change the number of needed driver nodes, but  may change the set of driver nodes. The critical edge attack strategy \cite{Sun2019ICSRS} collects all the critical edges from the initial network and remove them, thereafter a random edge attack is performed. This is more destructive than a simple degree- or betweenness-based attack in the early stage of the process.

In this paper, a hierarchical framework is proposed for both node- and edge-removal attacks, aiming at maximizing the destruction of the network controllability. The main contributions of this work are: 1) the concept of {\it critical node} is introduced, quantified and analyzed, as a complement to the concept of critical edge; 2) a new hierarchical attack framework is proposed, which sorts the destruction of nodes or edges in a descending order, and is updated iteratively; 3) extensive simulations are performed to verify the effectiveness of the proposed methods, revealing that the exposure of critical edges and nodes is harmful to maintain a good network controllability.

The rest of the paper is organized as follows. Section \ref{sec:pre} reviews the network controllability and its robustness, and several existing attack strategies. Section \ref{sec:haf} introduces a new hierarchical attack framework. Section \ref{sec:exp} evaluates both the hierarchical node- and edge-removal strategies by extensive numerical simulations, on both synthetic and real-world networks. Section \ref{sec:end} concludes the investigation.

\section{Preliminary}
\label{sec:pre}

\subsection{Controllability and Controllability Robustness}
\label{sub:cr}

A linear time-invariant (LTI) networked system, described by $\dot{{\bf x}}=A{\bf x}+B{\bf u}$, is \textit{state controllable} if and only if the controllability matrix $[B\ AB\ A^2B\ \cdots A^{N-1}B]$ has a full row-rank, where $A$ and $B$ are constant matrices of compatible dimensions, $\bf{x}$ is the state vector, $\bf{u}$ is the control input, and $N$ is the dimension of $A$. The \textit{structural controllability} is its slight generalization dealing with two parameterized matrices $A$ and $B$, in which the parameters characterize the structure of the underlying networked system: if there are specific parameter values that can ensure the system to be state controllable, then the system is structurally controllable. If the system is state controllable, its state vector $\bf{x}$ can be driven from any initial state to any target state in the state space within finite time by a suitable control input $\bf{u}$. Clearly, without control input $\bf{u}$, or $B\equiv 0$, the networked system is by no means controllable. Likewise, for a network of one-dimensional (scalar) nodes, there exist control inputs to some nodes to ensure its controllability. This network controllability is characterized by the minimum number of nodes with control inputs, called driver nodes, needed to maintain the controllability. When the network is put into the above LTI system formulation, how many and which nodes should be driver nodes are described by the matrix $B$.

Specifically, the controllability of a network of $N$ scalar nodes is measured by the density of the driver nodes $n_D$, defined by
\begin{equation}\label{eq:nd}
n_D\equiv \frac{N_D}{N}\,,
\end{equation}
where $N_D$ is the minimum number of driver nodes needed to retain the network controllability. Smaller $n_D$ value represents better controllability. Practically, $N_D$ can be calculated in two ways, for structural controllability and for exact (state) controllability, respectively. It was shown in \cite{Liu2011N} that identifying the minimum number of driver nodes to achieve a full control of a directed network requires searching for a maximum matching of the network, which quantifies the network structural controllability. When a maximum matching is found, $N_D$ is determined by the number of unmatched nodes, i.e., nodes without control inputs, given by
\begin{equation}\label{eq:sc}
N_D=\text{max}\{1, N-|E^*|\},
\end{equation}
where $|E^*|$ is the number of nodes in the maximum matching $E^*$. As for exact controllability \cite{Yuan2013NC}, $N_D$ is calculated by
\begin{equation}\label{eq:ec}
N_D=\text{max}\{1, N-\text{rank}(A)\}.
\end{equation}

The controllability robustness is evaluated after some nodes or edges are removed, one by one, yielding a sequence of values (represented by a \textit{controllability curve}) that reflect how robust (or vulnerable) a networked system is against a destructive attack. The controllability curve under a node-removal attack is calculated by
\begin{equation}\label{eq:ndin}
n_D^{N}(i)\equiv \frac{N_D(i)}{N-i}\,,\ \ i=0,1,\ldots,N-1,
\end{equation}
where $N_D(i)$ is the number of driver nodes needed to retain the network controllability after $i$ nodes have been removed, and $N$ represents the number of nodes in the original network. Note that, given an $N$-node network, one can remove at most $N-1$ nodes, excluding the trivial empty case. Similarly, the controllability curve under an edge-removal attack is calculated by
\begin{equation}\label{eq:ndie}
n_D^{E}(i)\equiv \frac{N_D(i)}{N}\,,\ \ i=0,1,\ldots,M,
\end{equation}
where $N_D(i)$ is the number of driver nodes needed to retain the network controllability after $i$ edges have been removed, and $N$ and $M$ represent the numbers of nodes and edges in the original network. Here, $n_D^{N}(0)=n_D^{E}(0)$ represents the controllability of the original network, of which no node or edge has been removed.

To measure the overall controllability robustness of a network, the controllability curves are averaged:
\begin{equation}\label{eq:rcn}
R^N_c= \frac{1}{N} \sum_{i=0}^{N-1}n_D^{N}(i),
\end{equation}
and
\begin{equation}\label{eq:rce}
R^E_c= \frac{1}{M+1} \sum_{i=0}^{M}n_D^{E}(i).
\end{equation}
Lower $R^N_c$ and $R^E_c$ values mean better overall controllability against node- and edge-removal attacks, respectively.

\subsection{Existing Attack Strategies}
\label{sub:eas}

The most frequently used measures of importance are the \textit{degree} and \textit{betweenness} centralities. A weighted measure is given by
\begin{equation}\label{eq:bi}
p_i=\alpha\times\frac{k_i}{\sum_{i=1}^{N}{k_i}} + \beta\times\frac{b_i}{\sum_{i=1}^{N}{b_i}}\,,
\end{equation}
where $k_i$ and $b_i$ represent the degree and the betweenness of node $i$, $p_i$ represents the probability of removing it, and $\alpha$ and $\beta$ are weights, which are set manually in \cite{Nie2015PA} with $\beta$ being replaced by $1-\alpha$ in \cite{Gao2018PA}. Similarly, in \cite{Hao2020PA} three parameters, $\alpha$, $\beta$ and $\gamma$, are used to control the weights of degree, betweenness and harmonic closeness, respectively. These measures have been used in the strategies to attack interdependent networks \cite{Huang2011PRE,Dong2012PRE,Gao2018PA,Cui2018PA,Hao2020PA}, networks of networks \cite{Dong2013PRE,Liu2015CSF}, and weighted networks \cite{Bellingeri2018PA}.

\begin{figure}[htbp]
	\centering
	\includegraphics[width=.9\linewidth]{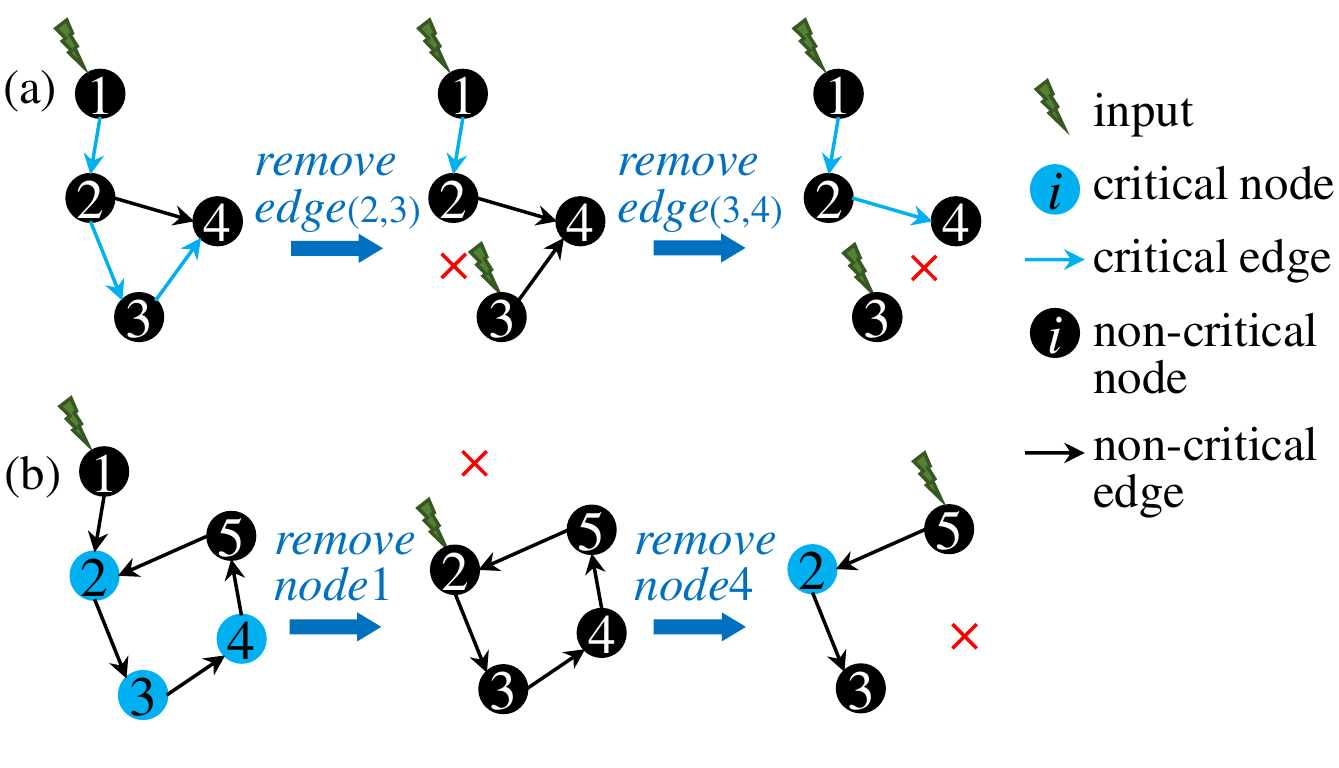}
	\caption{[color online] An example of critical edges and nodes changing during the attack process: (a) edge $(2,4)$ is non-critical in the initial network, but becomes critical after some edge-removals; (b) nodes $2$,$3$ and $4$ are critical initially, but node $1$ is removed, all of them become non-critical, and after node $4$ is removed, node $2$ becomes critical again.}
	\label{fig:change}
\end{figure}
The edge-removal attack strategy proposed in \cite{Sun2019ICSRS} aims at removing the critical edges of the initial network, and after all the initial critical (IC) edges have been removed a random attack is performed. This IC attack strategy is specifically designed to degrade the network controllability, where the term `critical' is defined for controllability. This attack is especially destructive in the early stage of the process, but becomes less effective in latter stages, because critical edges are changing during the process due to the removal of some other edges. An example is shown in Fig. \ref{fig:change} (a), where a non-critical edge (edge $(2,4)$) becomes critical after some edge-removals. Therefore, critical edges need to be updated throughout the attack process such that the damage to network controllability can be maximized.

\subsection{Critical Edges and Nodes}
\label{sub:cri}

In this paper, the concept of critical edge defined in \cite{Liu2011N} is adopted. An edge is critical if and only if its removal increases the number of driver nodes needed to retain the network controllability; otherwise, it is non-critical. Inspired by this, the concept of critical node is introduced here. A node is critical if and only if its removal increases the number of driver nodes needed to retain the network controllability; otherwise, it is non-critical. An example of critical node is shown in Fig. \ref{fig:change} (b), where the blue-colored nodes are critical nodes.

The critical nodes and edges are the most important elements in the concern of network controllability, in the sense that their removal will cause the maximum possible destruction to the network controllability. Therefore, in an efficient attack strategy, critical nodes and edges should be removed with the highest priority. It should be noted that, through the attack process, both critical nodes and edges will be dynamically changed, as illustrated by the example shown in Fig. \ref{fig:change}. Therefore, in analyzing the attack strategy and its effect, the list of critical nodes and edges must be updated iteratively, step by step, after each attack.

\section{Hierarchical Attack Framework}
\label{sec:haf}
\subsection{Hierarchical Edge Attack}
\label{sub:hea}

\begin{figure}[htbp]
	\centering
	\includegraphics[width=.9\linewidth]{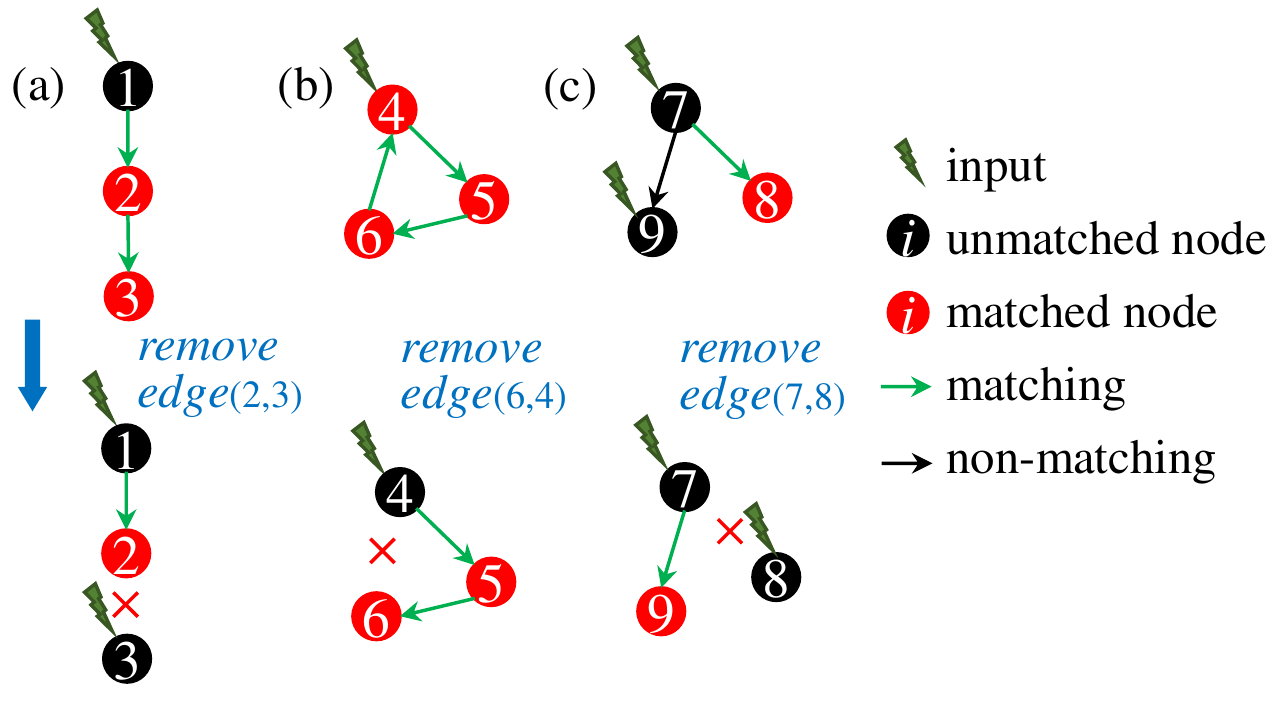}
	\caption{[color online] Example of edge hierarchy: (a) edge $(2,3)$ is critical, whose removal will increase the number of needed driver nodes by $1$; (b) edge $(4,6)$ is subcritical, whose removal will not change the number of driver nodes but will increase the number of unmatched nodes by $1$; (c) edge $(7,8)$ is normal, whose removal does  not change the numbers of driver nodes and unmatched nodes.}
	\label{fig:ecri}
\end{figure}

The proposed framework classifies all edges hierarchically into three types: 1) critical edges, as defined above; 2) subcritical edges, whose removal does not increase the number of needed driver nodes, but increases the number of unmatched nodes; and 3) normal edges, which are the rest edges. The subcritical and normal edges are non-critical edges. In a hierarchical attack, the priorities of attacking these three types are in descending order, namely, selecting the critical edges with the highest priority to attack, followed by the subcritical ones, and finally the normal ones. An example of these three types of edges is shown in Fig. \ref{fig:ecri}.

Algorithm \ref{alg:edge} shows the pseudo-code for hierarchical edge selection. Given a network with $M$ edges, represented by its adjacency matrix $A$, Algorithm \ref{alg:edge} returns the index of the edge to be removed with the highest priority. Lines $1$--$3$ initialize three empty lists for the three types of edges. Lines $4$ and $5$ calculate the numbers of needed driver nodes and unmatched nodes of the original network before being attacked. The for-loop between Lines $6$--$20$ categorizes each edge into a type list. In Lines $21$--$27$, the non-empty list with the highest priority is assigned to $L$, which is then sorted according to a certain feature $F$ (e.g., degree centrality). Finally, $L(1)$ represents the index of an edge that is with the highest priority to be removed, and meanwhile it has the highest value of feature $F$ (e.g., highest degree).

\subsection{Hierarchical Node Attack}
\label{sub:hna}

\begin{figure}[htbp]
	\centering
	\includegraphics[width=.9\linewidth]{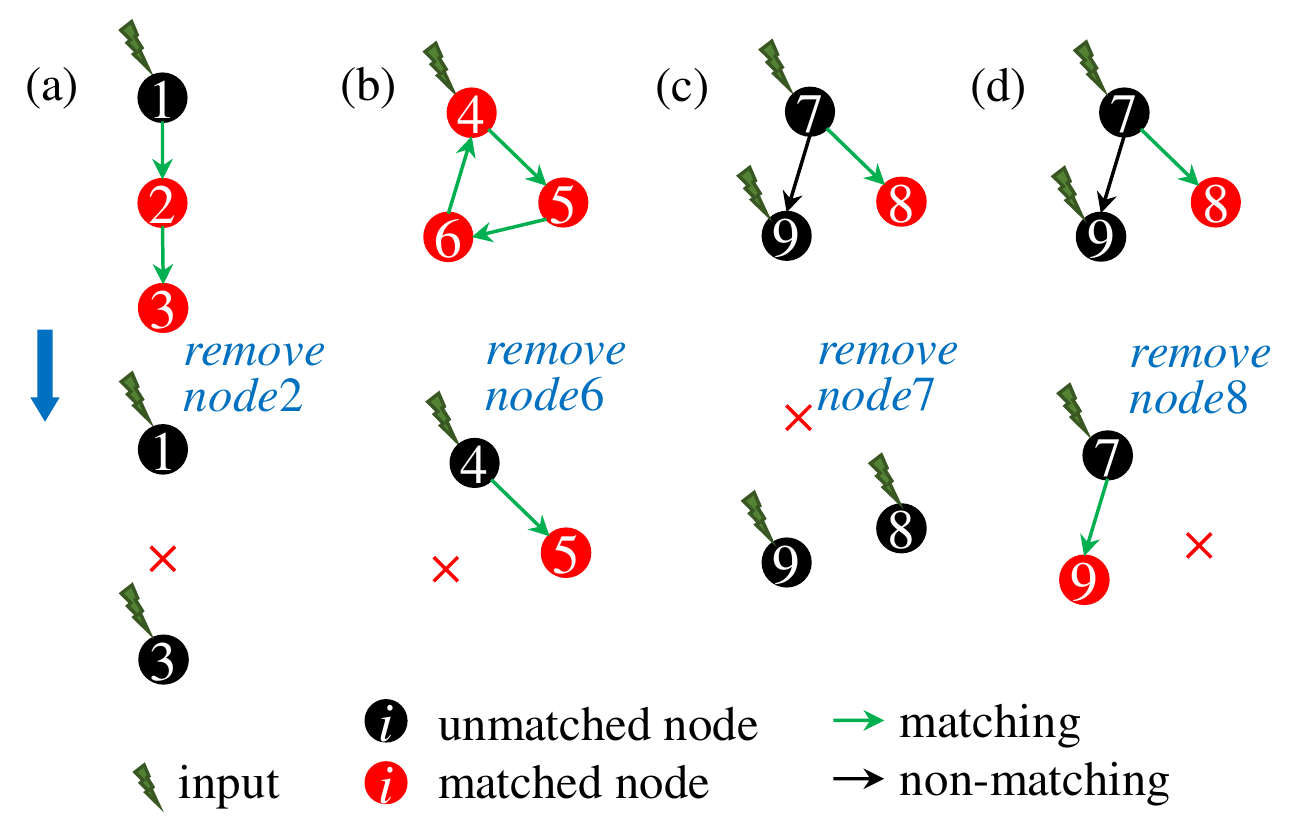}
	\caption{[color online] Example of node hierarchy: (a) node $2$ is critical, whose removal increases the number of needed driver nodes by $1$; (b) node $6$ is subcritical, whose removal does not change the number of driver nodes, but increases the number of unmatched nodes by $1$; (c) node $7$ is normal, whose removal does not change the numbers of driver nodes and unmatched nodes; (d) node $8$ is redundant, whose removal decreases the number of needed driver nodes by $1$.}
	\label{fig:ncri}
\end{figure}

Different from edges, nodes are hierarchically classified into four types in descending order of priorities: 1) critical nodes; 2) subcritical nodes, whose removal does not increase the number of needed driver nodes but increases the number of unmatched nodes; 3) normal nodes, whose removal does not affect the numbers of driver nodes and unmatched nodes; and 4) redundant nodes, whose removal enhances the controllability contrarily. The subcritical, normal, and redundant nodes are non-critical nodes. An example of these four types of nodes is shown in Fig. \ref{fig:ncri}.

Algorithm \ref{alg:node} shows the pseudo-code for hierarchical node selection. Given a network with $N$ nodes, represented by its adjacency matrix $A$, Algorithm \ref{alg:node} returns the index of the node to be removed with the highest priority. Lines $1$--$4$ initialize four empty lists for the four types of nodes. Lines $5$ and $6$ calculate the numbers of needed driver nodes and unmatched nodes of the original network before being attacked. The for-loop between Lines $7$--$23$ categorizes each node into a type list. In Lines $24$--$32$, the non-empty list with the highest priority is assigned to $L$, which is then sorted according to certain feature $F$. Finally, $L(1)$ represents the index of a node that has the highest priority to be removed, and meanwhile it has the highest value of feature $F$.

Source codes of both hierarchical node and edge attack algorithms are available for the public\footnote{\url{https://fylou.github.io/sourcecode.html}}.

\begin{algorithm}
	\SetKwData{Left}{left}\SetKwData{This}{this}\SetKwData{Up}{up}
	\SetKwFunction{Union}{Union}\SetKwFunction{FindCompress}{FindCompress}
	\SetKwInOut{Input}{Input}\SetKwInOut{Output}{Output}
	\SetAlgoLined
	\Input{adjacency matrix $A$; feature $F$; number of edges $M$}
	\Output{index $j$ of the edge to be attacked}
	{$L_1\leftarrow$ []; \tcp{highest priority}}
	{$L_2\leftarrow$ []\;}
	{$L_3\leftarrow$ []; \tcp{lowest priority}}
	{$n_A \leftarrow$ \emph{number of driver nodes needed for $A$}\;}
	{$u_A \leftarrow$ \emph{number of unmatched nodes needed for $A$}\;}
	\For{$i\leftarrow 1$ \KwTo $M$}
	{
		{$A_0\leftarrow A$\;}
		{\emph{Delete edge $i$ from $A_0$}\;}
		{$n_{A_0} \leftarrow$ \emph{number of driver nodes needed for $A_0$}\;}
		{$u_{A_0} \leftarrow$ \emph{number of unmatched nodes needed for $A_0$}\;}
		\eIf{$n_{A_0}>n_A$}
		{$L_1.insert(i)$\;}
		{\eIf{$u_{A_0}>u_A$}{$L_2.insert(i)$\;}{$L_3.insert(i)$\;}}
	}
	\uIf{$L_1$ is not empty}
	{$L\leftarrow L_1$\;}
	\uElseIf{$L_1$ is empty \textbf{and} $L_2$ is not empty}
	{$L\leftarrow L_2$\;}
	\Else
	{$L\leftarrow L_3$\;}
	{\emph{Sort $L$ according to feature $F$, in descending order}\;}
	{$j\leftarrow L(1)$\;}
	\caption{Hierarchical Edge Selection}\label{alg:edge}
\end{algorithm}

\begin{algorithm}
	\SetKwData{Left}{left}\SetKwData{This}{this}\SetKwData{Up}{up}
	\SetKwFunction{Union}{Union}\SetKwFunction{FindCompress}{FindCompress}
	\SetKwInOut{Input}{Input}\SetKwInOut{Output}{Output}
	\SetAlgoLined
	\Input{adjacency matrix $A$; feature $F$; number of nodes $N$}
	\Output{index $j$ of the node to be attacked}
	{$L_1\leftarrow$ []; \tcp{highest priority}}
	{$L_2\leftarrow$ []\;}
	{$L_3\leftarrow$ []\;}
	{$L_4\leftarrow$ []; \tcp{lowest priority}}
	{$n_A \leftarrow$ \emph{number of driver nodes needed for $A$}\;}
	{$u_A \leftarrow$ \emph{number of unmatched nodes needed for $A$}\;}
	\For{$i\leftarrow 1$ \KwTo $N$}
	{
		{$A_0\leftarrow A$\;}
		{\emph{Delete node $i$ from $A_0$}\;}
		{$n_{A_0} \leftarrow$ \emph{number of driver nodes needed for $A_0$}\;}
		{$u_{A_0} \leftarrow$ \emph{number of unmatched nodes needed for $A_0$}\;}
		\uIf{$n_{A_0}>n_A$}
		{$L_1.insert(i)$\;}
		\uElseIf{$n_{A_0}=n_A$}
		{\eIf{$u_{A_0}>u_A$}{$L_2.insert(i)$\;}{$L_3.insert(i)$\;}}
		\Else{$L_4.insert(i)$\;}
	}
	\uIf{$L1$ is not empty}
	{$L\leftarrow L_1$\;}
	\uElseIf{$L1$ is empty \textbf{and} $L_2$ is not empty}
	{$L\leftarrow L_2$\;}
	\uElseIf{$L_1$, $L_2$ are empty \textbf{and} $L_3$ is not empty}
	{$L\leftarrow L_3$\;}
	\Else
	{$L\leftarrow L_4$\;}
	{\emph{Sort $L$ according to feature $F$, in descending order}\;}
	{$j\leftarrow L(1)$\;}
	\caption{Hierarchical Node Selection}\label{alg:node}
\end{algorithm}

\subsection{Extra Computational Complexity}
\label{sub:ecc}

The computational complexity of calculating the network controllability, mainly in searching for the number of needed driver nodes, is $O(M\cdot\sqrt{N})$, by the Hopcroft--Karp algorithm. In a hierarchical node or edge attack, to identify whether a node or edge is critical, the number of needed driver nodes to be calculated iteratively introduces a non-negligible amount of extra computational cost. This extra computational cost for hierarchical edge attack is $O(\sum_{i=1}^{M}(i\cdot\sqrt{N}))$, and for hierarchical node attack is $O(\sum_{i=1}^{N-1}(i\cdot\sqrt{N}))$.

\section{Experimental Studies}
\label{sec:exp}

In this section, the hierarchical attack framework is evaluated by extensive simulations. Network features will be taken into account. For node attacks, betweenness, out-degree and closeness are used as feature $F$, respectively; for edge attacks, betweenness and degree are used, respectively. To verify the effectiveness of the proposed hierarchical framework, the hierarchical feature-based attack strategies are compared to the feature-based attack strategies, respectively. For example, the hierarchical degree-based attack is compared to the degree-based attack, under the same conditions.

Nine typical directed synthetic network models are adopted for simulation, namely the Erd{\"{o}}s--R{\'e}nyi random-graph (ER) network \cite{Erdos1964RG}, Newman--Watts small-world (SW) network \cite{Newman1999PLA}, generic scale-free (SF) network \cite{Pu2012PA,Goh2001PRL,Sorrentino2007CH}, \textit{q}-snapback (QS) network \cite{Lou2018TCASI},
\textit{q}-snapback network with redirected edges (QR) \cite{Lou2019R}, random triangle (RT) network \cite{Chen2019TCASII}, and random rectangle (RR) network \cite{Chen2019TCASII}, extremely homogeneous (HO) network \cite{Lou2020TCASI}, and onion-like (OL) network \cite{Herrmann2011JSTAT}.

Recall that the HO networks were empirically verified with optimal controllability robustness before \cite{Lou2020TCASI}. Given an $N$-node and $M$-edge configuration, the HO network satisfies $\floor{M/N}\leq k_{i}^{in,out}\leq\ceil{M/N}$, $i=1,2,\ldots,N$. This means that both of its in- and out-degrees are distributed identically or nearly identically with a small difference less than $1$. The OL network is generated from an SF network via simple edge-swapping with degree reservation \cite{Herrmann2011JSTAT}, thus its degree distribution follows the same power-law distribution as the SF.

The detailed generation methods and parameter settings of these synthetic networks are provided in Supplementary Information (SI)\footnote{\url{https://fylou.github.io/pdf/hatk_si.pdf}}. The network size is set to $N=500$, $1000$, and $1500$, respectively. The average degree is set to $\langle k\rangle=3$, $5$, and $10$, respectively.

In addition, nine real-world networks are used for simulations, with data taken from Network Repository\footnote{\url{http://networkrepository.com/}}. Their parameters and brief descriptions are presented in Table \ref{tab:rwn_para}.

\begin{table}[htbp]
	\centering
	\caption{Basic information of the real-world networks.}
	\begin{tabular}{|c|l|l|c|c|}
		\hline
		\begin{tabular}[c]{@{}c@{}}Network\end{tabular}& \multicolumn{1}{c|}{File name}&\multicolumn{1}{c|}{Brief description} &$N$&$M$ \\ \hline
		BMK&\begin{tabular}[c]{@{}l@{}}bn-mouse-kasthuri\\-graph-v4\end{tabular}&brain network&1029&1559 \\ \hline
		ICM&ia-crime-moreno&interaction network&830&1474 \\ \hline
		IEU&inf-euroroad&infrastructure network &1175&1417 \\ \hline
		DEL&delaunay-n10&DIMACS10 problem&1024&3056 \\ \hline
		DW5&dwt-1005&\multirow{2}{*}{\begin{tabular}[c]{@{}l@{}}symmetric connection\\ from Washington\end{tabular}}&1005&3808 \\ \cline{1-2} \cline{4-5}
		DW7&dwt-1007& &1007&3784 \\ \hline
		LSH&lshp1009&\begin{tabular}[c]{@{}l@{}}Alan George’s\\L-shape problem\end{tabular}       &1009&2928 \\ \hline
		OLM&olm1000&\begin{tabular}[c]{@{}l@{}}computational fluid\\dynamics problem\end{tabular}  &1000&2996 \\ \hline
		RAJ& rajat19&\begin{tabular}[c]{@{}l@{}}Rajat19 circuit\\simulation matrix\end{tabular}     &1157&4433 \\ \hline
	\end{tabular}\label{tab:rwn_para}
\end{table}

\subsection{Comparison of Attack Strategies}
\label{sub:cas}

\subsubsection{Node-removal attacks}

Nine node-removal attack strategies are compared, namely the betweenness-based (B), out-degree-based (D), closeness-based (C), random (R), hierarchical betweenness-based (HB), hierarchical out-degree-based (HD), hierarchical closeness-based (HC), hierarchical random (HR), and hybrid (Hy) attacks.

The B, D, C strategies aim at removing the node with the largest betweenness, degree, and closeness, respectively, at every step. The HB, HD, HC strategies aim at removing critical nodes that are sorted in a betweenness-, degree-, and closeness-descending order, respectively, at every step. The HR removes the critical nodes at random.

The Hy strategy is designed as follows: First, remove either the node (or edge) with the maximum degree (or betweenness), according to the removal of which node (or edge) will cause greater destruction to the network controllability. If equal, then choose either one to attack.

\subsubsection{Edge-removal attacks}

Eight edge attack strategies are compared, namely the betweenness-based (B), out-degree-based (D), random (R), hierarchical betweenness-based (HB), hierarchical out-degree-based (HD), hierarchical random (HR), initial critical (IC) \cite{Sun2019ICSRS} and hybrid (Hy) attacks.

Here, the `out-in' edge degree is used, which is defined as the sum of the out-degree of its source node and the in-degree of its target node \cite{Sun2019ICSRS}.

For each node and each edge attack, the simulation repeats $30$ and $20$ independent runs, respectively.

\subsection{Simulation Results on Synthetic Networks}
\label{sub:expsn}

Here, the structural controllability (see Eq. (\ref{eq:sc})) is considered for controllability robustness comparison. The simulation results of some synthetic networks with $N=1000$ and real-world networks are presented. More detailed and complete results for networks with $N=500$, $1000$ and $1500$ are given in the SI.

\begin{figure}[htbp]
	\centering
	\includegraphics[width=\linewidth]{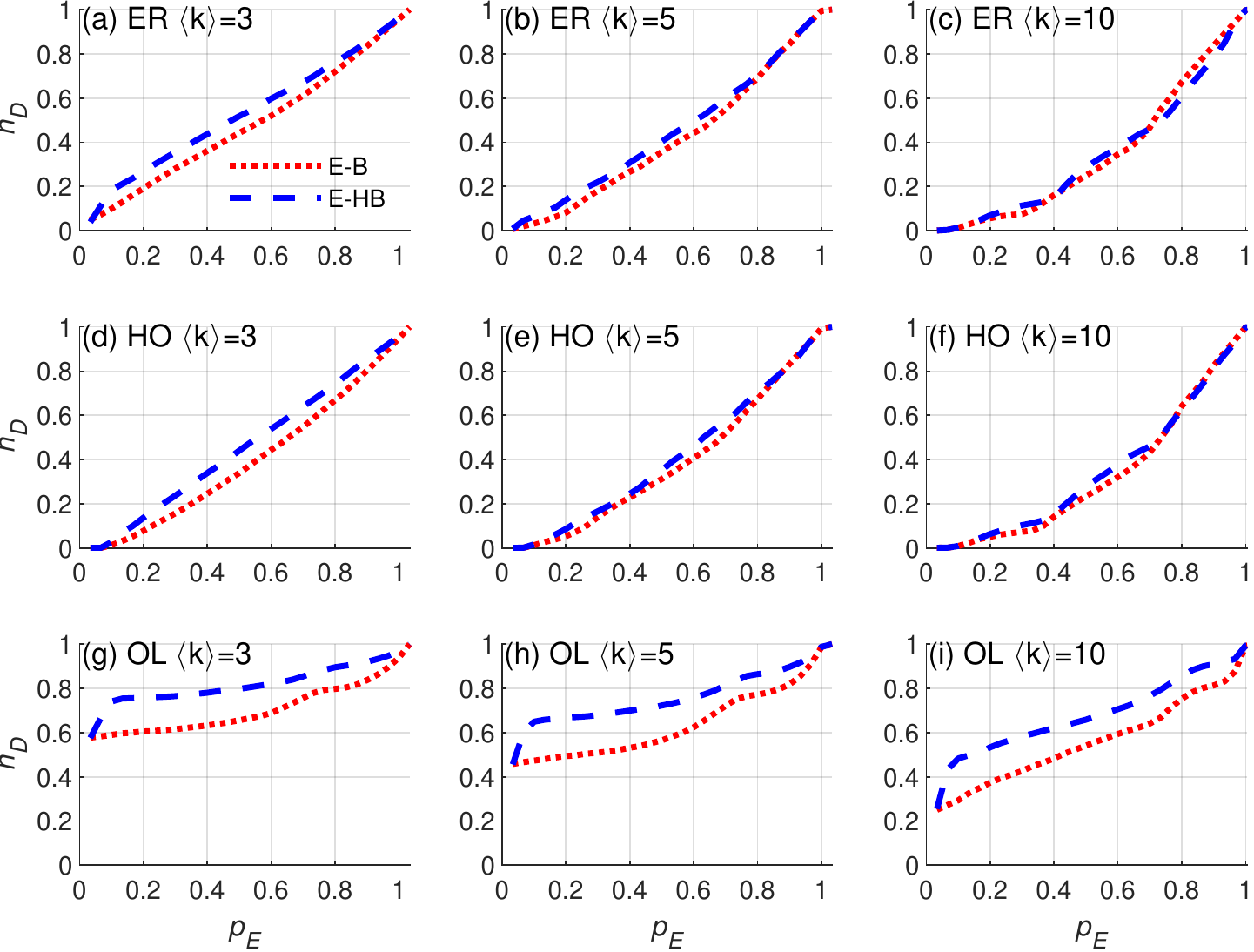}
	\caption{[color online] Results of edge attacks on ER, HO, and OL ($N=1000$): hierarchical betweenness-based (E-HB) and betweenness-based (E-B) strategies.}
	\label{fig:hbvsb}
\end{figure}

Fig. \ref{fig:hbvsb} shows the results of ER, HO, and OL under E-HB and E-B attacks. It is clear that E-HB is consistently more destructive than E-B throughout the entire process as shown in Figs. \ref{fig:hbvsb} (a,b,d,e,f,g,h,i); while Fig. \ref{fig:hbvsb} (c) shows that E-HB is more destructive than E-B when $P_E<0.7$, but E-B is slightly more destructive when $P_E>0.7$.

\begin{figure}[htbp]
	\centering
	\includegraphics[width=\linewidth]{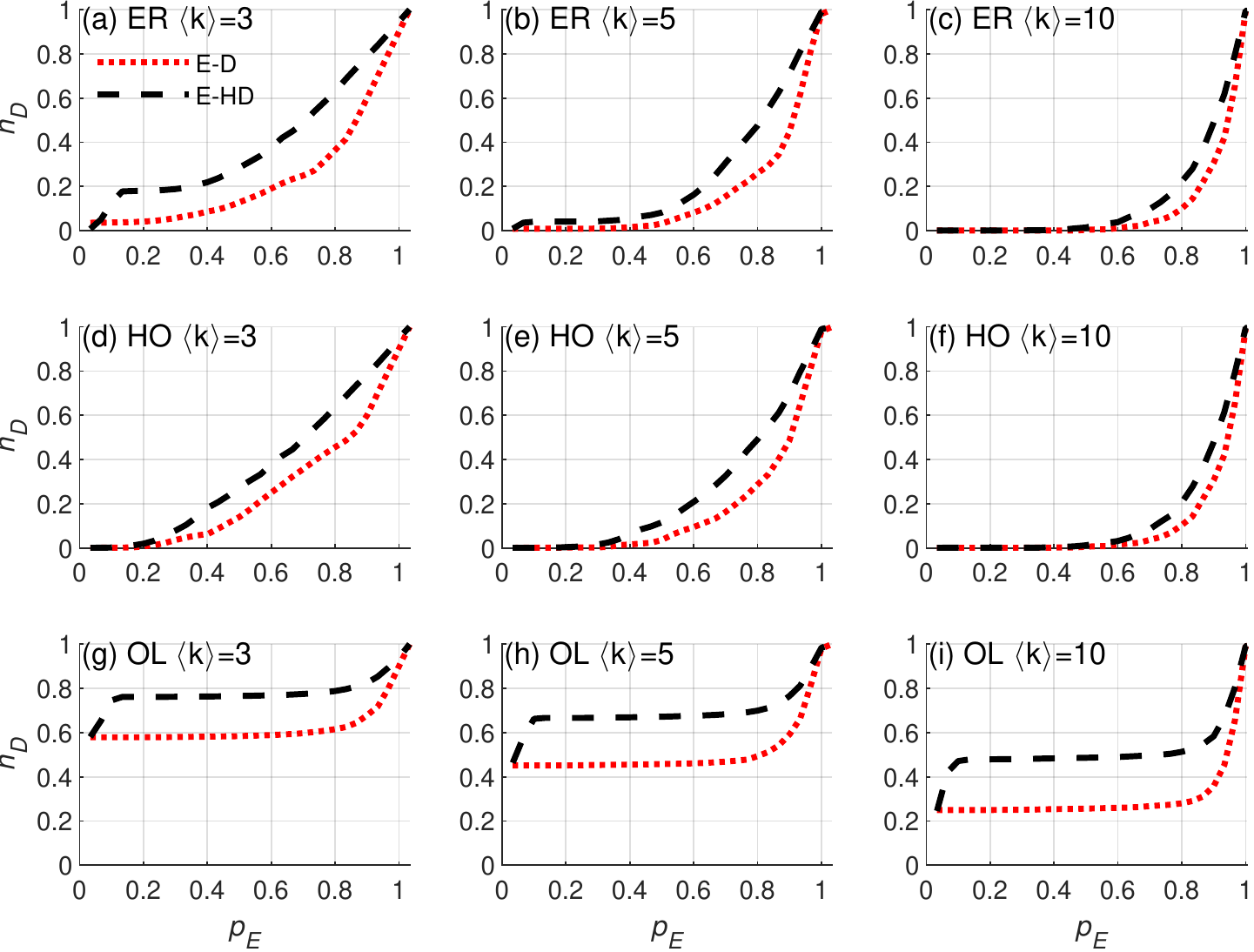}
	\caption{[color online] Results of edge attacks on ER, HO, and OL ($N=1000$): hierarchical degree-based (E-HD) and degree-based (E-D) strategies.}
	\label{fig:hdvsd}
\end{figure}

\begin{figure}[htbp]
	\centering
	\includegraphics[width=.95\linewidth]{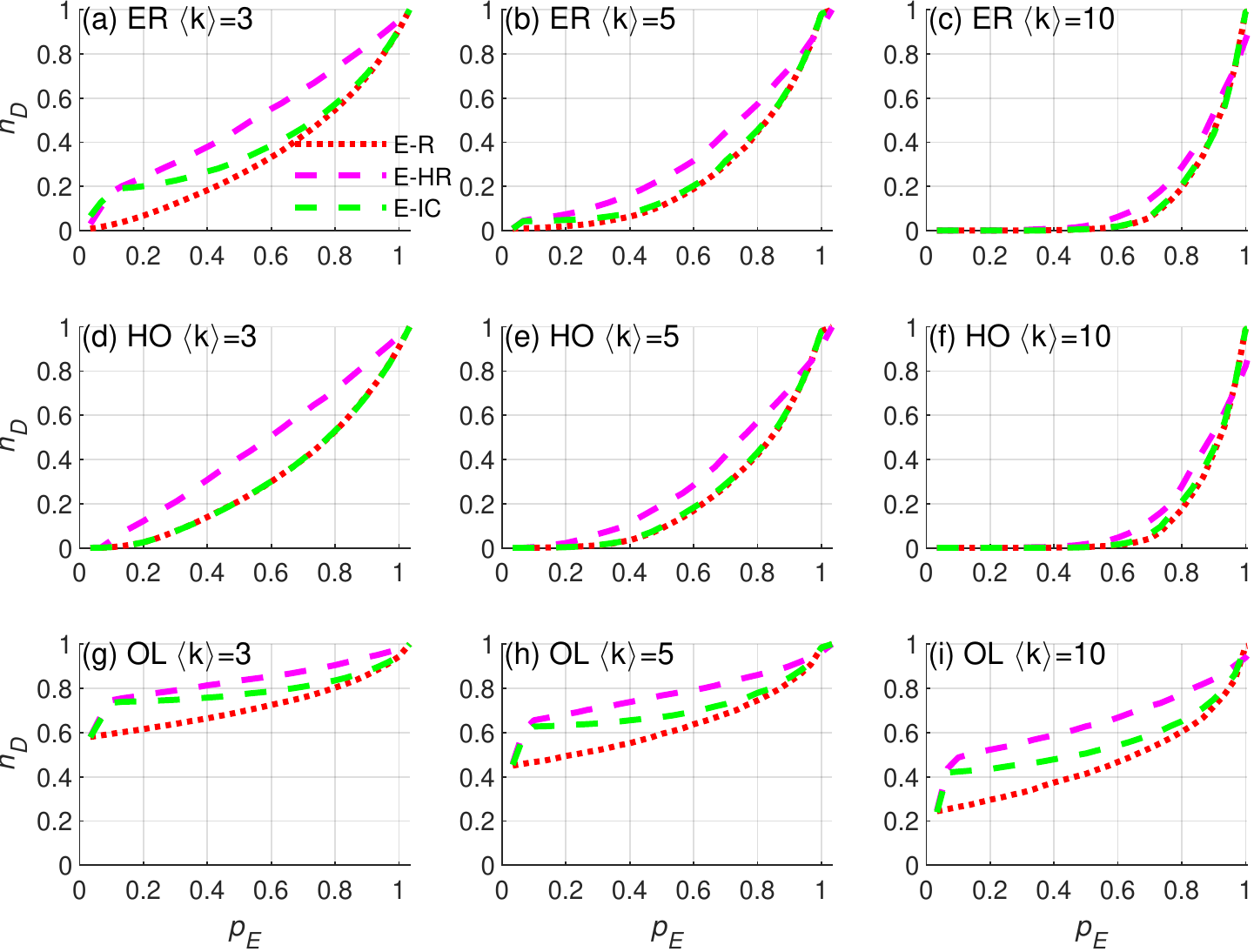}
	\caption{[color online] Results of edge attacks on ER, HO, and OL ($N=1000$): random (E-R), hierarchical random (E-HR) and initial critical (IC) strategies.}
	\label{fig:hrvsr}
\end{figure}

Fig. \ref{fig:hdvsd} shows that E-HD is consistently more destructive than E-D. Fig. \ref{fig:hrvsr} shows that E-HR is consistently more destructive than E-R and E-IC. Figs \ref{fig:hbvsb}--\ref{fig:hrvsr} show that HO has better controllability robustness than ER; both HO and ER have significantly better controllability robustness than OL. As the average degree increases, the controllability robustness improves.

\begin{figure}[htbp]
	\centering
	\includegraphics[width=.95\linewidth]{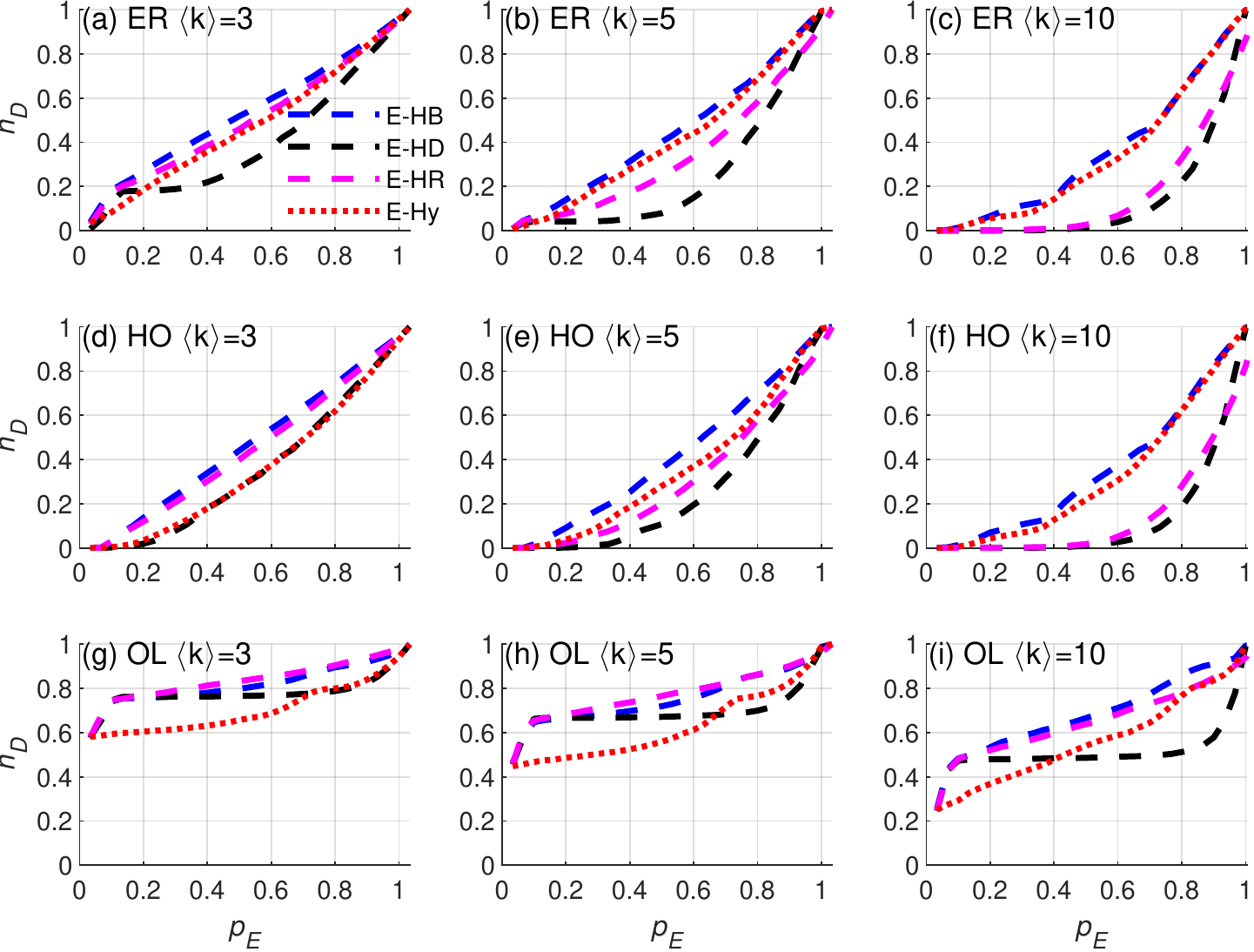}
	\caption{[color online] Results of edge attacks on ER, HO, and OL ($N=1000$): three hierarchical attacks (E-HB, E-HD and E-HR) and hybrid (E-Hy) strategies.}
	\label{fig:hyb}
\end{figure}

Fig. \ref{fig:hyb} compares the three hierarchical attacks (E-HB, E-HD and E-HR) and hybrid attack (E-Hy). For ER and HO, it is clear that E-HB is more destructive than the other strategies; while for OL, either E-HR or E-HB is most destructive. Figs. \ref{fig:hbvsb}--\ref{fig:hyb} show that the hierarchical framework increases the destruction effects on the network controllability. It can also be seen that, among the hierarchical attack strategies, E-HB performs the best.

The overall comparison is summarized in Table \ref{tab:n1000}, where each value is the ratio of the overall destruction (see Eq. (\ref{eq:rcn}) for node attacks and Eq. (\ref{eq:rce}) for edge attacks) under the two corresponding attack strategies. For example, the value $1.210$ in row `$\langle k\rangle=3$, ER' and column `Node Attack, HB/B' represents that, given an ER ($\langle k\rangle=3$) under node attacks, the overall destruction ratio of N-HB versus N-B is $1.210$. Referring to Fig. \ref{fig:hbvsb} (a), the value is equivalent to the ratio of the area under the blue dashed-line versus the area under the red dotted-line.

If HB/B $>1$, it means that HB is more destructive; if HB/B $<1$, it means that B is more destructive; otherwise, if HB/H $=1$, it means that HB and B are equivalently destructive. Here, an equivalent overall destruction does not mean that the two controllability curves are overlapped, but means that the areas under the two controllability curves are equal, namely they are equivalent in the average sense.

As can be seen from the `Node Attacks' part in Table \ref{tab:n1000}, the hierarchical attack strategies are consistently more destructive than the non-hierarchical and the hybrid strategies, for the ratio values are greater than $1$ in the columns of HB/B, HD/D, HC/C, and HR/R.

In the columns of HB/Hy and HD/Hy, hierarchical strategies generally outperform the hybrid strategy. It should be noted that, when the ratio is within $[0.990,1.010]$, one may consider the two comparing strategies to have equivalent performances.

As for the edge-removal attacks, the hierarchical strategies are more destructive than the non-hierarchical ones and the IC (which does not update the critical edge list). However, the ratio values in the column of HD/Hy are mostly less than $1$, meaning that HD is less destructive than Hy. This implies that edge degree is not a good measure of importance regarding destructive attacks. Nevertheless, within the hierarchical framework, E-HD is more destructive than E-D.

The results for cases of $N=500$ and $N=1500$ are tabled in SI.

\begin{table*}[htbp]
	\centering
	\caption{Comparison of attack strategies on the nine synthetic networks ($N=1000$), where B represents betweenness; D represents degree; C represents closeness; R represents random; Hy represents hybrid; IC represents initial critical edges; HB represents hierarchical betweenness; HD represents hierarchical degree; HC represents hierarchical closeness; HR represents hierarchical random attacks.}
	\begin{tabular}{|c|c|c|c|c|c|c|c|c|c|c|c|c|c|}
		\hline
		\multicolumn{2}{|c|}{\multirow{2}{*}{N=1000}}&\multicolumn{6}{c|}{Node Attack}&\multicolumn{6}{c|}{Edge Attack}              \\ \cline{3-14}
		\multicolumn{2}{|c|}{}&HB/B&HD/D&HC/C&HR/R&HB/Hy&HD/Hy&HB/B&HD/D&HR/R&HB/Hy&HD/Hy&HR/IC \\ \hline
		\multirow{9}{*}{$\langle k\rangle$=3}&ER&1.210&1.151&1.313&1.469&1.111&1.132&1.112&1.585&1.454&1.123&0.848&1.225 \\ \cline{2-14}
		&SW&1.286&1.070&1.209&1.275&1.065&1.081&1.157&1.339&1.426&1.295&0.997&1.426 \\ \cline{2-14}
		&SF&1.033&1.011&1.029&1.266&1.007&1.009&1.158&1.246&1.157&1.157&1.101&1.060 \\ \cline{2-14}
		&QS&1.303&1.198&1.661&1.344&1.098&1.185&1.123&2.256&1.463&1.123&0.886&1.241 \\ \cline{2-14}
		&QR&1.336&1.160&1.330&1.445&1.138&1.155&1.118&1.721&1.440&1.122&0.874&1.411 \\ \cline{2-14}
		&RT&1.148&1.095&1.164&1.590&1.056&1.070&1.260&1.679&1.393&1.280&1.072&1.202 \\ \cline{2-14}
		&RR&1.245&1.110&1.240&1.527&1.095&1.098&1.217&1.682&1.438&1.231&1.013&1.282 \\ \cline{2-14}
		&HO&1.247&1.170&1.436&1.416&1.141&1.148&1.151&1.336&1.378&1.095&1.143&1.382 \\ \cline{2-14}
		&OL&1.034&1.012&1.028&1.273&1.007&1.007&1.161&1.241&1.161&1.164&1.104&1.060 \\ \hline
		\multirow{9}{*}{$\langle k\rangle$=5}&ER&1.207&1.209&1.339&1.483&1.149&1.165&1.051&1.377&1.354&1.052&0.627&1.272 \\ \cline{2-14}
		&SW&1.296&1.171&1.337&1.383&1.135&1.150&1.067&1.421&1.261&1.142&0.756&1.292 \\ \cline{2-14}
		&SF&1.051&1.017&1.048&1.371&1.015&1.015&1.179&1.348&1.208&1.191&1.101&1.084 \\ \cline{2-14}
		&QS&1.235&1.268&1.984&1.348&1.167&1.214&1.066&2.097&1.476&1.054&0.732&1.477 \\ \cline{2-14}
		&QR&1.273&1.197&1.345&1.502&1.152&1.156&1.078&1.416&1.366&1.077&0.673&1.370 \\ \cline{2-14}
		&RT&1.201&1.152&1.222&1.630&1.110&1.118&1.106&1.543&1.378&1.176&0.751&1.339 \\ \cline{2-14}
		&RR&1.239&1.154&1.264&1.545&1.109&1.120&1.111&1.435&1.305&1.132&0.756&1.326 \\ \cline{2-14}
		&HO&1.234&1.176&1.399&1.431&1.150&1.136&1.074&1.413&1.258&1.026&0.994&1.246 \\ \cline{2-14}
		&OL&1.048&1.019&1.046&1.367&1.010&1.010&1.205&1.340&1.199&1.199&1.098&1.077 \\ \hline
		\multirow{9}{*}{$\langle k\rangle$=10}&ER&1.214&1.208&1.320&1.450&1.148&1.117&1.025&1.325&1.202&1.030&0.470&1.221 \\ \cline{2-14}
		&SW&1.217&1.208&1.363&1.376&1.149&1.113&1.064&1.238&1.127&1.102&0.479&1.164 \\ \cline{2-14}
		&SF&1.078&1.036&1.080&1.590&1.025&1.027&1.205&1.608&1.336&1.197&0.916&1.155 \\ \cline{2-14}
		&QS&1.167&1.315&2.875&1.295&1.198&1.195&1.036&1.777&1.395&1.188&0.576&1.369 \\ \cline{2-14}
		&QR&1.228&1.203&1.359&1.427&1.148&1.116&1.029&1.341&1.154&1.044&0.487&1.137 \\ \cline{2-14}
		&RT&1.211&1.208&1.286&1.496&1.121&1.128&1.075&1.332&1.280&1.057&0.515&1.221 \\ \cline{2-14}
		&RR&1.224&1.169&1.300&1.409&1.106&1.119&1.090&1.309&1.214&1.083&0.555&1.187 \\ \cline{2-14}
		&HO&1.201&1.194&1.361&1.376&1.157&1.140&1.034&1.275&1.063&1.078&0.697&1.063 \\ \cline{2-14}
		&OL&1.066&1.034&1.079&1.577&1.021&1.022&1.189&1.619&1.369&1.212&0.932&1.167 \\ \hline
	\end{tabular}\label{tab:n1000}
\end{table*}

The attack simulation results on real-world networks are shown in Table \ref{tab:rwn}, while the detailed comparison figures for different networks are included in SI.

The attack simulations on various synthetic networks and real-world networks show that the hierarchical strategies are consistently more destructive to network controllability than other attack strategies.

\begin{table*}[htbp]
	\centering
	\caption{Comparison of attack strategies on the nine real-world networks, where B represents betweenness; D represents degree; C represents closeness; R represents random; Hy represents hybrid; IC represents initial critical edges; HB represents hierarchical betweenness; HD represents hierarchical degree; HC represents hierarchical closeness; HR represents hierarchical random attacks.}
	\begin{tabular}{|c|c|c|c|c|c|c|c|c|c|c|c|c|}
		\hline
		\multirow{2}{*}{}&\multicolumn{6}{c|}{Node Attack}             &\multicolumn{6}{c|}{Edge Attack}              \\ \cline{2-13}
		& HB/B &HD/D &HC/C &HR/R &HB/Hy&HD/Hy&HB/B &HD/D &HR/R &HB/Hy&HD/Hy&HR/IC \\ \hline
		BMK              &1.066&1.007&1.010&1.157&1.004&1.005&1.031&1.063&1.056&1.030&0.988&1.027 \\ \hline
		ICM              &1.044&1.102&1.183&1.361&1.033&1.041&1.101&1.201&1.187&1.100&1.004&1.088 \\ \hline
		IEU              &1.187&1.050&1.137&1.361&1.007&1.040&1.109&1.377&1.291&1.115&1.097&1.096 \\ \hline
		DEL              &1.203&1.174&1.191&1.311&1.064&1.057&1.108&1.416&1.248&1.112&0.889&1.163 \\ \hline
		DW5              &1.276&1.149&1.456&1.423&1.097&0.995&1.167&1.483&1.381&1.169&0.838&1.237 \\ \hline
		DW7              &1.323&1.036&1.597&1.314&1.004&0.994&1.205&2.224&1.284&1.235&1.291&1.269 \\ \hline
		LSH              &1.301&1.312&1.977&1.264&1.066&1.086&1.121&1.585&1.301&1.122&0.907&1.217 \\ \hline
		OLM              &1.007&1.001&1.009&1.946&1.010&1.019&1.281&1.660&1.611&1.696&1.627&1.649 \\ \hline
		RAJ              &1.297&1.076&1.252&1.769&1.064&1.082&1.508&1.749&1.502&1.805&1.403&1.292 \\ \hline
	\end{tabular}\label{tab:rwn}
\end{table*}

\subsection{Critical Edges and Nodes}
\label{sub:cen}

A common phenomenon is observed from the results presented in Sec. \ref{sub:expsn}: as the average degree increases, the ratio of areas under the controllability curves subject to hierarchical and non-hierarchical attacks tends asymptotically to $1$. This is because, as the network becomes denser, hence more homogeneous, fewer critical edges and nodes are exposed. It not only improves the controllability robustness of the network, but also makes the proposed hierarchical attack strategies less effective, thereby becoming similar to non-hierarchical attacks.

\begin{table}[htbp]
	\centering
	\caption{The lowest average out-degree when there is no critical nodes or edges found in the network.}
	\begin{tabular}{|c|c|c|c|c|c|c|c|c|c|}
		\hline
		&ER&SW&SF&QS&QR&RT&RR&HO&OL \\ \hline
		Node&9&3& 23&6&4&8&6&3&22 \\ \hline
		Edge&10&3&24&5&6&8&6&3&22 \\ \hline
	\end{tabular}\label{tab:noncri}
\end{table}

Table \ref{tab:noncri} shows the minimum (integer) average degree when there is no critical nodes or edges found in the initial network. Here, initial means the network that has not been attacked. For each topology with a given $\langle k\rangle$ value, $30$ network instances are simulated. Given $N=500$, $\langle k\rangle$ is set from $3$ with an incremental value $1$. If there are no critical nodes or edges found in all the $30$ instances, then the $\langle k\rangle$ value is recorded into Table \ref{tab:noncri}; otherwise, $\langle k\rangle$ increases by $1$ and then the process is run again. It can be seen from the table that, for SW and HO, there are no critical nodes or edges found when $\langle k\rangle=3$, meaning that removal of any node or edge in the initial SW or HO will not increase the number of needed driver nodes. Thus, their controllability robustness is better than the others. In contrast, for SF and OL, before $\langle k\rangle$ increases to $22$ and $24$, respectively, there were no critical nodes or edges. It means that in dense SF or OL networks (e.g., $\langle k\rangle=20$), there are still critical nodes and edges, and removing any critical node or edge will directly destroy its controllability. Thus, SF and OL have much worse initial controllability and controllability robustness than the other networks.

\begin{figure}[htbp]
	\centering
	\includegraphics[width=.9\linewidth]{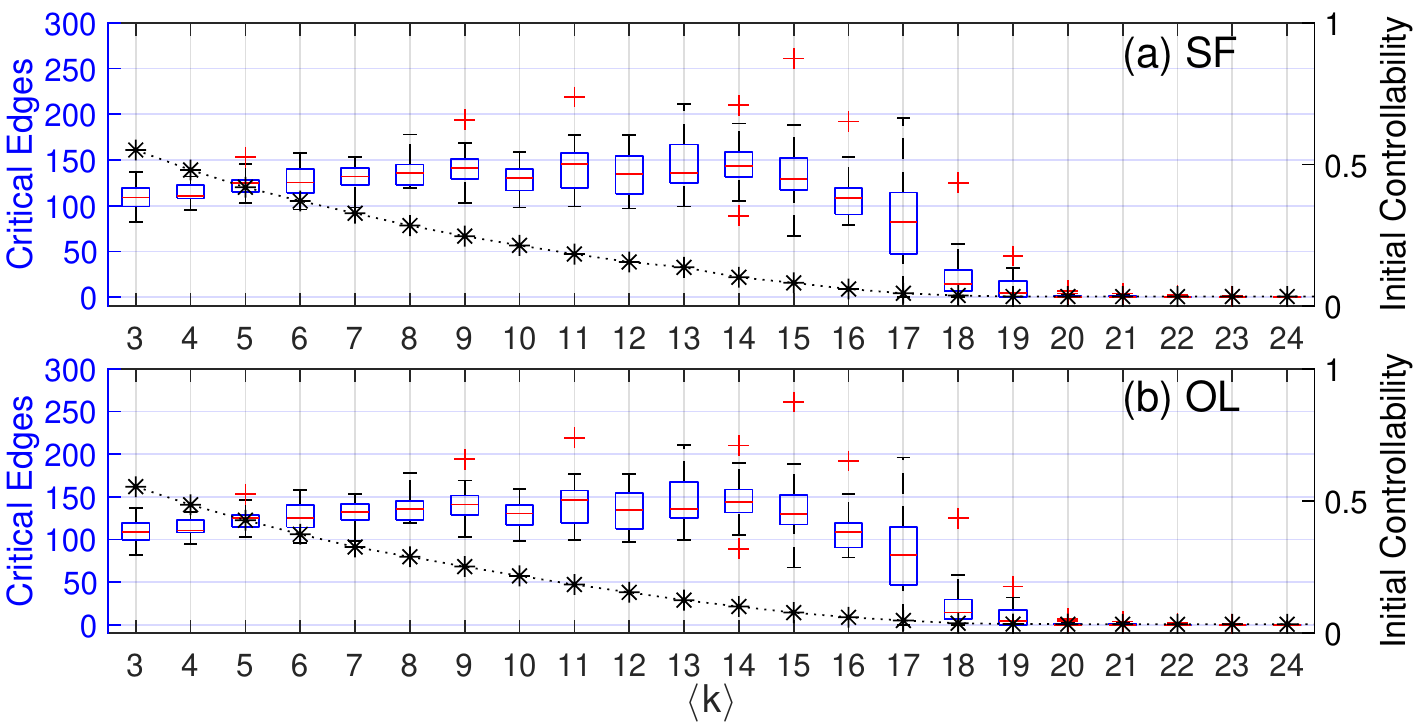}
	\caption{Number of critical edges (boxplots) and initial controllability (stars *) against the average degree of (a) SF and (b) OL networks.}
	\label{fig:cri_edge}
\end{figure}

Fig. \ref{fig:cri_edge} shows the number of critical edges in the initial SF and OL networks, against the increase of average degree. The corresponding figure for the case with critical nodes is shown in SI. The initial controllability is plotted for reference in each subplot. As shown in Fig. \ref{fig:cri_edge}, the numbers of critical edges in the initial SF and OL networks increase as $\langle k\rangle$ increases from $3$ to $13$; when $\langle k\rangle>16$, the numbers of critical edges drop drastically. Meanwhile, the initial controllability of both SF and OL becomes better as the average degree increases. When $3\leq\langle k\rangle\leq13$, the additional edges enhance the connectedness and make the networks more controllable. These additional edges become part of the critical edges. However, when $\langle k\rangle>16$, the initial controllability of the SF networks tends to be sufficiently optimized, reflected by the lower density of needed driver nodes. In this case, the increased edges cover the critical nodes and edges, which leads to the drastically drops of the numbers of (the exposed) critical edges.

The exposure of critical nodes and edges sets a clear target for the attacker to destroy the network controllability. In contrast, in the networks with strong controllability robustness, there are rare (or no) critical nodes and edges exposed; for example, SW, HO, QS and QR. For these networks, the attacker is unable or uneasy to find targets to attack in order to destruct the controllability. This finding is consistent with, and actually extends the applicability of, the previous findings: 1) dense and homogeneous networks have better controllability \cite{Liu2011N}; 2) extremely-homogeneous topology has the optimal controllability robustness \cite{Lou2020TCASI}. Nevertheless, critical nodes and edges will expose themselves during the attack process, as the network becomes sparser. To design networks with good controllability robustness, the exposure of critical nodes and edges should be dimmed or avoided, if ever possible. If there are sufficient numbers of available edges, the networks should be designed as dense and homogeneous as possible \cite{Liu2011N,Lou2020TCASI}; otherwise, if the numbers of edges are limited, they should be deliberately assigned in such a way that the exposure of critical nodes and edges is minimum.

\begin{figure*}[htbp]
	\centering
	\includegraphics[width=.90\linewidth]{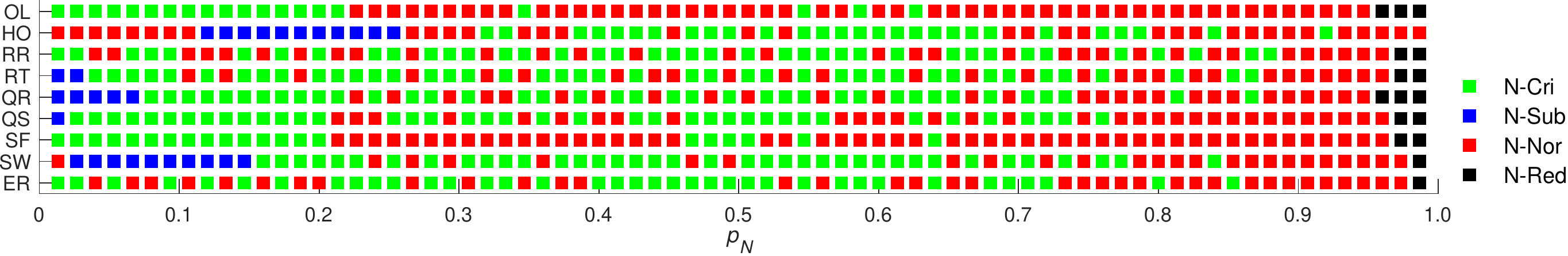}
	\caption{[color online] Types of the removed nodes under N-HR attacks. The network configuration is $N=1000$ and $\langle k\rangle=5$.}
	\label{fig:type}
\end{figure*}

Fig. \ref{fig:type} shows the types of removed nodes and edges during an attack. As can be seen from Fig. \ref{fig:type} (a), HO and SW do not expose critical nodes in the early stage, and thus the attacker could only remove the normal or subcritical nodes; later, in the middle stage, both HO and SW expose some critical nodes; finally, SW has only redundant nodes left. The other $7$ networks expose more critical nodes than HO or SW during the attack process. Note that, although for OL and SF, there are more normal nodes than critical ones during the attack process, their initial controllability is no good (see the controllability curves in SI). 

\section{Conclusions}
\label{sec:end}

To better understand the network controllability robustness from the perspective of destructive attacks, a hierarchical attack framework is proposed, which can be used for both edge- and node-removal attacks. The hierarchical attack strategies aim at removing the critical nodes and edges with the highest priority, and they can be combined together with other commonly used features (e.g., degree or edge centrality), such that the identified critical nodes or edges can be sorted in descending order according to such features. Extensive experiments on nine synthetic networks with various topologies and nine real-world networks show the effectiveness of the proposed hierarchical attack framework on destructive attacks to network controllability for all kinds of networks that are tested. For node attacks, betweenness, out-degree, and closeness are used as the feature, respectively; for edge attacks, betweenness and degree are used, respectively. The hierarchical feature-related attacks show consistently better destructive performances than the common feature-only attacks.

It is revealed that the exposure of the critical edges and nodes are disadvantageous in resisting attacks to the network controllability. Therefore, to design networks with strong controllability robustness, the critical nodes and edges should be deliberately hidden. This finding is consistent with, and also extends the applicability of, the previous findings: 1) dense and homogeneous networks have better controllability \cite{Liu2011N}; 2) extremely-homogeneous topology has the optimal controllability robustness with fixed numbers of nodes and edges \cite{Lou2020TCASI}.

%
%
%



\end{document}